# EuFe$_2$As$_2$ magnetic structure determination

Javier Herrero-Martín[1]*, Valerio Scagnoli[1], Claudio Mazzoli[1], Yixi Su[2], Ranjan Mittal[2,3], Yinguo Xiao[4], Thomas Brueckel[2,4], Neeraj Kumar[5], S. K. Dhar[5], A. Thamizhavel[5] and Luigi Paolasini[1]

[1] European Synchrotron Radiation Facility, BP 220, F-38043 Grenoble Cedex 9, France

[2] Juelich Centre for Neutron Science, IFF, Forschungszentrum Juelich, Outstation at FRM II, Lichtenbergstrasse 1, D-85747 Garching, Germany

[3] Solid State Physics Division, Bhabha Atomic Research Centre, Trombay, Mumbai 400 085, India

[4] Institut fuer Festkoerperforschung, Forschungszentrum Juelich, D-52425 Juelich, Germany

[5] Department of Condensed Matter Physics and Material Sciences, Tata Institute of Fundamental Research, Homi Bhabha Road, Colaba, Mumbai 400 005, India

Magnetic resonant x-ray scattering experiments have been performed on a single crystal of EuFe$_2$As$_2$ at the Eu L$_3$ absorption edge. The orientation of Eu magnetic moments was directly determined: in the antiferromagnetic ordered phase they lay parallel to the crystallographic a-axis. In addition, non-resonant magnetic x-ray measurements indicate that Fe magnetic moments are aligned along the same direction in the spin density wave phase. As deduced by temperature dependence of integrated intensities, the Fe magnetic arrangement seems to be insensitive to the onset of Eu AFM phase. Some speculations on the low temperature space group are reported on the base of detected resonant reflections on forbidden Bragg positions.

# 1. INTRODUCTION

Very recently reported high $T_C$ superconductivity in the electron/hole-doped $R$FeAsO ($R$=La-Gd) ("1111") [1-3] and $A$Fe$_2$As$_2$ ($A$: Ca, Sr, Ba, and Eu) ("122") [4-7] families of iron pnictide compounds has been one of the most striking discoveries of the last years in condensed matter physics. The structure of the iron pnictide compounds is tetragonal at room temperature (*I4/mmm*), where the buffer layer of $R$-O/$A$ atoms and the superconductivity relevant Fe-As layer alternate along the c-axis. In undoped parent compounds, a tetragonal to orthorhombic structural transition takes place at low temperatures. This transition is usually accompanied by a concomitant antiferromagnetic ordering of small Fe moments, which has been interpreted as due to spin density wave (SDW) ordering of iron [8-11]. While the mechanism of superconductivity in iron pnictides is still far from clear, it has been established that the emergence of superconductivity and the suppression of static magnetic ordering can be effectively achieved via either electron or hole doping from suitable chemical substitutions at the *R/A* or Fe sites [4,8,9]. Different from the case in cuprates, rather surprisingly, superconductivity may coexist with the static magnetism in the underdoped regime of certain iron pnictide compounds [12-15]. This highlights the importance of the interplay between superconductivity and magnetism.

EuFe$_2$As$_2$ is the only known "122" compound containing magnetic rare earth ion in the *A*-site. Therefore it is an ideal example to study the interplay between the magnetism from Eu and Fe, due to the presence of a large localized magnetic moment of Eu$^{2+}$ ions (~7 $\mu_B$), which orders below $T_N$ ~ 19 K in a likely A-type antiferromagnetic (AFM) structure, as suggested from recent magnetic susceptibility and specific heat measurements [16]. A structural transition occurs at $T_S$ ~ 190 K lowering the symmetry from tetragonal to orthorhombic *Fmmm* (a=5.55 Å, b=5.50 Å and c=12.06 Å at 10 K) [10] cell. In this paper, reflections are indexed using the orthorhombic notation. The change in the crystal structure is accompanied by a SDW ordering of Fe spins. Bulk measurements suggest that the Fe magnetic moments are aligned within the ab-plane [17]. In fact, it has been determined via neutron diffraction that the Fe magnetic moments in both "1111" and "122" iron pnictide parent compounds are aligned along the crystal a-axis [3,18].

In EuFe$_2$As$_2$, the determination of the magnetic structure of both sublatttices via neutron diffraction appears not trivial due to the presence of Eu, a strong neutron absorber.

Therefore, non-resonant (NRXS) and element-specific resonant x-ray scattering (RXS) represents an appealing alternative to precisely determine the magnetic structure of $EuFe_2As_2$ [19].

The following analysis clearly indicates that europium and iron magnetic moments are all oriented parallel to the crystal a-axis as shown in the top inset of Figure 1. The temperature dependence of relevant reflections supports a picture where the onset of AFM ordering of Eu ions at $T_N$ barely affects the SDW on Fe, in agreement with previous reports pointing to a weak interaction between the two magnetic order parameters [20]. Finally, we report on the observation of some reflections resonant at the Eu $L_3$ edge and forbidden in the low temperature space group.

## 2. EXPERIMENTAL DETAILS

Large single crystals of $EuFe_2As_2$ with typical dimensions $6 \times 4 \times 1mm^3$ were grown by the flux method, with starting composition Eu:Fe:As:Sn in the ratio 1:2:2:19, where Sn was later removed by centrifugation. Crystals were thoroughly characterized by chemical composition analysis via EDAX, resistivity, magnetization and specific heat measurements (Fig.1) and pre-oriented via x-ray Laue back reflection method.

Preliminary polarized neutron diffraction experiments were carried out on a ~50 mg single crystal at DNS at FRM II (Garching, Germany). The neutron wavelength was chosen at 4.74 Å. The spin-flip scattering of magnetic reflections was measured with a guide field perpendicular to the ($h$, 0, $l$) and (0, $k$, $l$) scattering planes as a function of temperature. This permitted to determine the magnetic propagation vector of both Fe SDW and Eu AFM structures as explained in the following.

X- ray scattering data were collected on ID20 beamline at ESRF (Grenoble, France) by using both horizontal ($\pi$ incident light) and vertical ($\sigma$ incident light) diffraction configurations [21]. A cleaved sample with the c-axis direction normal to the scattering surface was mounted in a refrigerator having a base temperature of ~10 K. The a-axis direction defines the azimuth reference throughout the paper.

The beamline optics were optimized close to Eu $L_3$ absorption edge (6.97 keV) for resonant measurements and to 6.93 keV for non-resonant one. In both cases, the scattered beam polarization was analyzed by means of a Cu (110) single crystal.

## 3. RESULTS AND DISCUSSION

Figure 2 shows reciprocal space contour maps measured at $T<T_N$ (a) and $T_N<T<T_S$ (b) by neutron diffraction. (001) and (003) reflections exist only at the lowest temperature. (103) reflection is not allowed in the low temperature orthorhombic space group (Fmmm). In figure 3, and by using the same experimental technique, the temperature dependences of some characteristic reflections are plotted. In particular, (003) reflection is only present below $T_N$, while (103) appears below $T_S$, suggesting that Eu and Fe spins arrangements are characterized by (001) and (101) reciprocal space vectors, respectively. An allowed Bragg reflection, (004), is also shown as reference for intensities evolution versus temperature.

In contrast to recently reported observations on the related compound $EuRh_2As_2$ [22], in our case reciprocal lattice investigation revealed magnetic orderings to be characterized by commensurate propagation vectors.

Once the propagation vectors known, we decided to exploit x-ray magnetic scattering technique for an easy determination of magnetic moment direction in the structure.

For Bragg forbidden reflections, x-ray atomic scattering factor can be written as

$$f = f' + if'' + f_{mag} \qquad (1)$$

where *f'* and *f''* account for dispersive and absorptive scattering processes, respectively, dominating the signal close to absorption edges, while $f_{mag}$ for non-resonant magnetic one.

At 6.93 keV, sufficiently far from any resonance in our compound, only $f_{mag}$ is expected to contribute. Moreover, at a temperature between $T_N$ and $T_S$ only Fe ions possess a long range ordered magnetic moment. According to neutron investigation, a x-ray intensity due to magnetic scattering is then expected in reciprocal lattice position characterized by (101) propagation vector. Below $T_N$ Eu ions order and a strong magnetic resonant signal around 6.97 keV should appear on reciprocal lattice position of (001) propagation vector. A series of magnetic reflections belonging to these two families where investigated. In the following we concentrate our attention on a subset of each family, all the other reflections bringing essentially the same information on the system.

A rocking curve of (109) magnetic reflection in non-resonant condition is shown in the top right inset of Figure 4. Resonant signals are characterized by similar profiles. Now on, by intensity of a reflection we mean the numerical integration over a theta scan (rocking curve) of detector counts divided by monitor ones. A linear estimation of the background is also subtracted.

Figure 3 also shows x-rays temperature dependences for a direct comparison with neutrons. In particular, we report (109) reflection off-resonance (not element specific) and (009) on-resonance (element specific) after correction by estimated beam heating. As expected, x-rays data follow the neutron ones, highlighting the correct attribution of the two reflections as due to Fe and Eu magnetic ordering, respectively. We believe that the small residual discrepancies with respect to neutron transition temperatures do not affect our magnetic structure analysis. Analogously to neutrons, (109) Fe magnetic reflection intensity does not change across $T_N$, probing the Fe order parameter to be insensitive to the Eu one. This fact supports a picture of two well decoupled magnetic substructures, in agreement with LSDA calculations [20].

Figure 4, shows an energy scan of the (109) magnetic reflection. The top left inset shows intensity variation versus azimuth angle collected at 6.93 keV and 30 K (symbols). Measurements were limited to the rotated σ–π' polarization channel, where the signal-to-noise ration is significantly better than in the un-rotated σ–σ' one. To account for absorption effects, azimuth experimental data are normalized by the permitted nuclear (2 0 10) reflection, lying close in the reciprocal space. The same correction has been applied to (209) reflection (see below).

The continuous line is a simulation performed by considering Fe moments aligned along the a-axis, resulting in a structure factor [23, 24]

$$F^a_{\sigma-\pi'}(109) \propto A + \cos(\psi) \qquad (2)$$

where $\psi$ is the azimuth angle and $A$ a constant factor determined by the experimental geometry. Below $T_S$ x-ray magnetic scattering indicates that Fe moments are aligned along the crystallographic a-axis.

The existence of Eu $L_3$ edge (E ~ 6.975 keV) resonance, although it is not related to the magnetic structure determination, deserves consideration.

A long range ordering of Fe magnetic moments is expected to polarize Eu 5d band, possibly resulting in a non-zero resonant intensity for reflections of suitable propagation vector, as already shown in other systems [25]. Of course in this case no signal is expected on the (019) reciprocal lattice position, not corresponding to any detected magnetic propagation vector in the compound. While no trace of non resonant signal was found at any of the inspected azimuth values on (019), below $T_S$ the reflection appears only in the rotated polarization channel and in resonant conditions. In Figure 4 we report also the measured energy spectrum of the (019) reflection for comparison. To explain the presence of this signal, a so-called Templeton scattering or anisotropic tensor of susceptibility (ATS) contribution has to be considered [26]. However, below $T_S$ the currently accepted space group is symmorphic, preventing the existence of any ATS signal [27, 28]. So, in our opinion it is likely that the *Fmmm* space group has to be revised. Nevertheless, (109) and (019) resonant intensities are extremely weak, comparable to the non-resonant magnetic signal from the Fe atoms. Therefore, deviations from the proposed structure, if any, should be very small and unlikely detectable by conventional powder and single-crystal diffraction techniques.

Now we turn our attention to the determination of Eu magnetic moments direction below $T_N$. We decided to study the (209) magnetic reflection to avoid the domain problems afflicting specular reflections (h=k=0).

Figure 5(a) shows the spectra of the (209) reflection in the σ-π' channel at 10 K in the vicinity of the Eu $L_3$ edge. The spectrum is characterized by a strong resonance centered at 6.973 keV, i.e. about 2-3 eV beyond the Eu $L_3$ absorption edge (see fluorescence). At resonance, as expected, there is no component in the σ-σ' channel [29] while the rotated polarization channel intensity is ~ 500 times stronger than the non resonant signal reported in Figure 4.

In order to determine the orientation of Eu magnetic moments below $T_N$ we have performed an azimuth angle scan of the (209) magnetic reflection.

Following [29], resonant structure factors for the (209) reflection in the rotated channel are

$$F^a_{\sigma-\pi'}(209) \propto A' + \cos(\psi) \quad (3)$$

$$F^b_{\sigma-\pi'}(209) \propto \sin(\psi) \quad (4)$$

for Eu moments aligned parallel to the a- or b-axis, respectively. A' is a constant factor analogous to A.

In the inset of Figure 5 measured resonant intensity in the σ-π' channel as a function of the azimuth angle is plotted together with the expected dependences derived from (3) and (4). The experimental data suggest that the Eu moments are also aligned along the crystallographic a-axis.

In order to independently confirm experimental evidences derived from azimuth analysis of the (209) reflection, below $T_N$ we have measured resonant intensities of several magnetic peaks in both π−π' and π−σ' channels, at two different values of azimuth angle.

Fig. 5(b) shows the ratio of $I_{\pi-\pi'}$ to $I_{\pi-\sigma'}$ for several magnetic reflections. Once again the model with Eu magnetic moments aligned along the a-axis better describes experimental data.

## 4. CONCLUSIONS

In summary, magnetism of iron pnictides is strongly related to their very interesting conductive capabilities, as demonstrated by correlation between Fe SDW and superconductivity in doped compounds. In this sense, the large magnetic moment on Eu is an ideal testing probe for possible interplay of Eu and (Fe, As) magnetic orderings.

In this paper, we have presented a neutron and x-ray magnetic scattering study of the magnetic structure of EuFe$_2$As$_2$: our experimental data suggest the Eu moments to be aligned along the crystallographic a-axis in the A-type AFM ordered phase. Moreover, we show that Fe moments in the SDW phase are aligned parallel to the a-axis as well. We have also observed that the onset of Eu magnetic ordering does not induce any noticeable change in Fe magnetic substructure.

We also report on the existence of Eu resonances above $T_N$ at Fe magnetic propagation vector and on (019) reflection. The origin of these signals is still not clear, possibly being related to structural effects.

**Acknowledgements**

The authors are grateful to P. Ghigna and F. de Bergevin for stimulating discussions, the ESRF for granting beam time, and A. Fondacaro for technical support.

**Figure captions**

Figure 1. Molar heat capacity of EuFe$_2$As$_2$ between 2 and 300 K. The structural ($T_S$ = 190 K) and magnetic ($T_N$ = 19 K) transitions are indicated. The inset shows the proposed magnetic structure for this material for T < $T_N$. The magnetic propagation vector defining the A-type AFM arrangement of Eu moments at is (001) and that of Fe is (101).

Figure 2. (Color online) Contour maps in the reciprocal space a-c plane at (a) T=10 K and (b) T=40 K obtained by neutron scattering. Miller indexes of nuclear (crystallographic) and magnetic reflections are labeled in black and white, respectively.

Figure 3. (Color online) Temperature dependence of selected magnetic reflections collected by x-rays and polarized neutron scattering. Left: (009) at the main resonance (filled triangles) and (109) at 6.930 keV (circles) around $T_N$. Neutron data collected on (003) reflection are represented by open triangles. Right: (109) at 6.930 keV around $T_S$ (filled circles) by x-rays compared to (103) (open circles) and (004) (crosses) as collected by neutron scattering in a wide T range from 2 to 220 K.

Figure 4. (Color online) Energy dependence of (109) (filled circles) and (019) (open circles) reflections in the σ-π' polarization channel at 30 K. The measured azimuth evolution of (109) at 6.930 keV (triangles) is shown in the left inset together with the calculated evolution (solid line) for Fe moments oriented parallel to the crystal a-axis. The right inset shows a rocking curve of the (109) reflection at the same energy and ψ = 170 deg.

Figure 5. (a) (Color online) Energy dependence of the σ-π' channel of (209) reflection (filled circles) at 10 K. Fluorescence is shown for comparison (crosses). The inset shows the experimental azimuth behavior of the (209) reflection (circles) compared to calculated evolution in case of Eu moments along crystal a- (solid line) or b- (dashed) axis. (b) Experimental intensity ratio ($I_{\pi\pi'}/I_{\pi\sigma'}$) for different resonant magnetic reflection associated to Eu order (filled circles) compared to the theoretical values for Eu moments aligned along a- (open circles) or b-axis (triangles). Reflections were collected with ψ ~45° and ψ~ 90°.

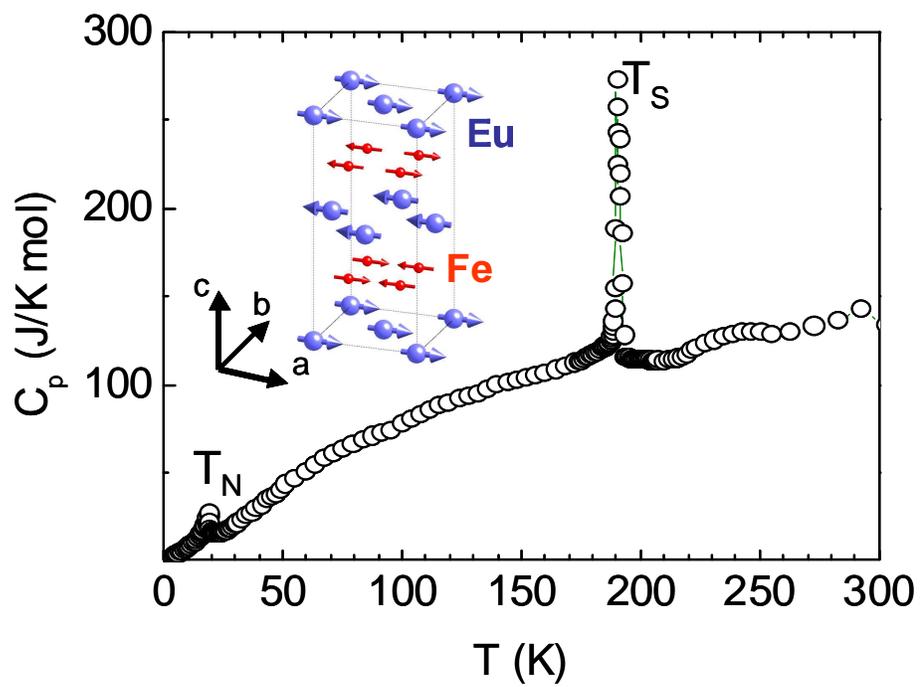

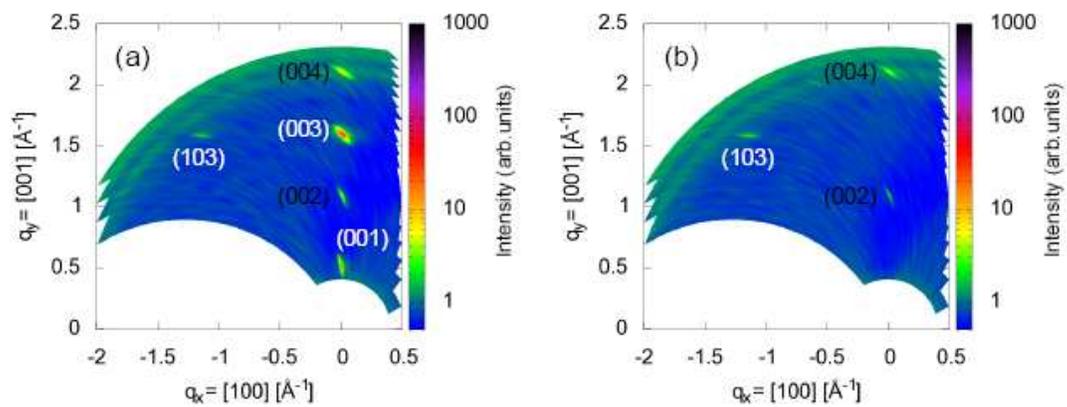

*Figure 2*

*Figure 3*

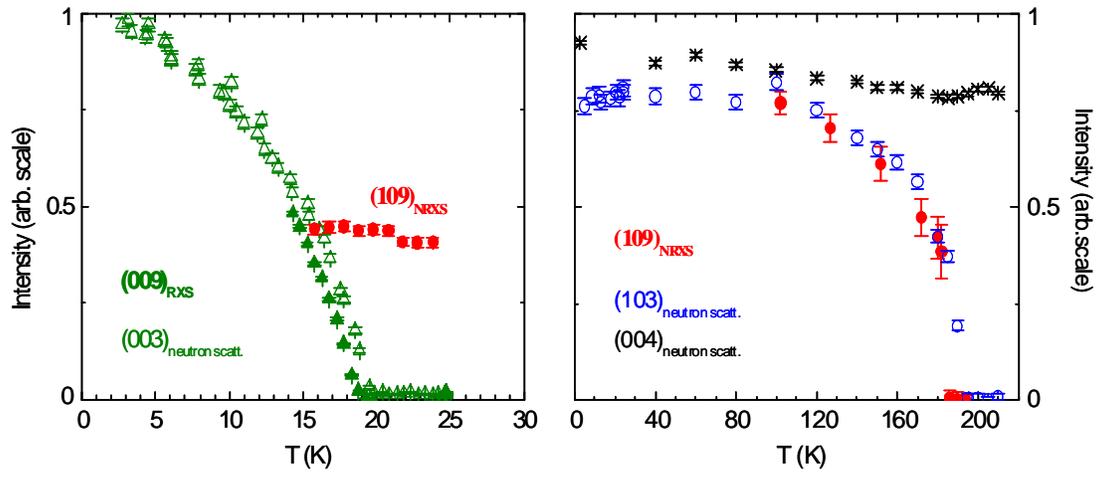

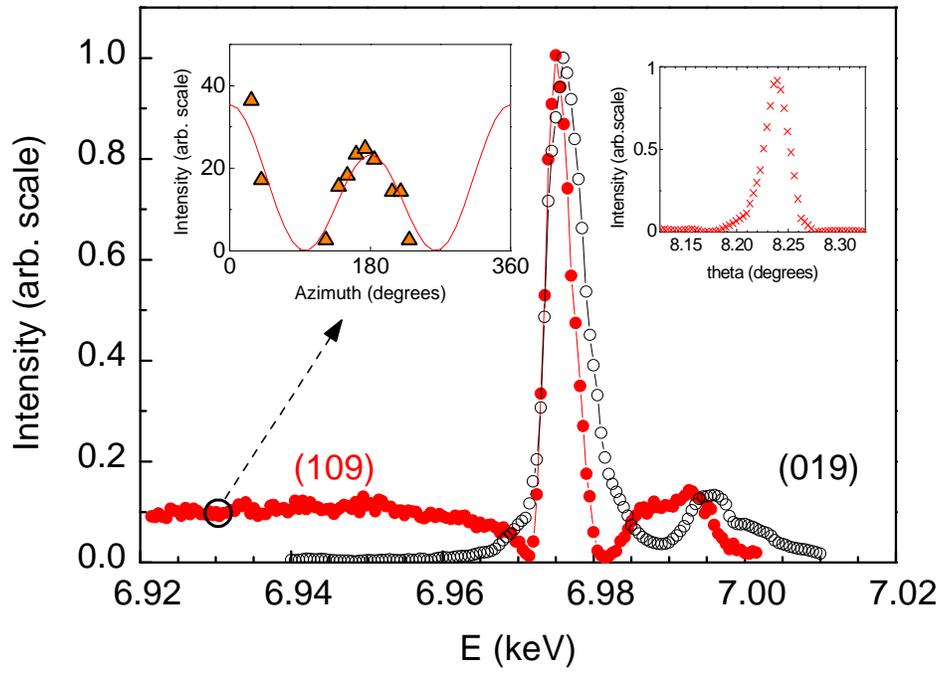

*Figure 5*

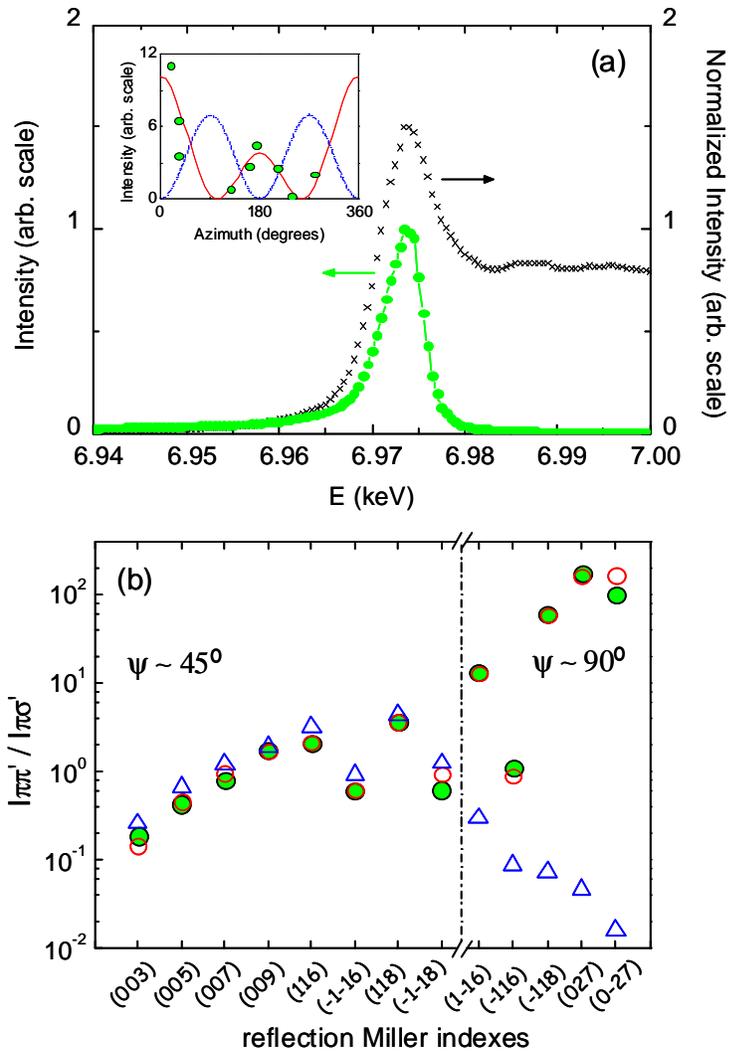